\documentstyle[prl,amsfonts,twocolumn,aps,epsfig]{revtex}
\catcode`\@=11 
\newcommand{\be}{\begin{equation}}
\newcommand{\ee}{\end{equation}}
\newcommand{\bea}{\begin{eqnarray}}
\newcommand{\eea}{\end{eqnarray}}

\newcommand{\p}{\partial}
\newcommand{\s}{\sigma}

\newcommand{\ri}{\mbox{i}}

\begin {document}
\title{
Restoration of Symmetry by Interactions and Nonreliability of the
Perturbative Renormalization Group Approach }
\author{ P. Azaria$~^{1}$,  P. Lecheminant$~^{2}$,  and A. M.
Tsvelik$~^{3}$}

\vspace{0.5cm}

\address{$^1$
Laboratoire de Physique Th{\'e}orique des Liquides,\\
Universit{\'e} Pierre et Marie Curie, 4 Place Jussieu, 75252 Paris,
France\\
$^2$ Laboratoire de Physique Th{\'e}orique et Mod{\'e}lisation,\\
Universit{\'e} de Cergy-Pontoise, Site de Saint Martin,
2 rue Adolphe Chauvin, 95302
Cergy-Pontoise Cedex, France\\
$^3$ Department of Physics, University of Oxford, 1 Keble Road,
Oxford OX1 3NP, UK}

\vspace{3cm}

\address{\rm (Received: )}
\address{\mbox{ }}
\address{\parbox{14cm}{\rm \mbox{ }\mbox{ }
 We discuss examples of 
(1+1)-dimensional models
where the perturbative renormalization group (RG)
indicates a tendency to restore the symmetry in the strong
coupling limit. We show that such restoration does occur sometimes, 
but the perturbative RG cannot be reliably used to detect it. 
}}
\maketitle
\makeatletter
\global\@specialpagefalse
\makeatother

Enlarged symmetry phenomenon at a fixed point corresponds to a situation 
where the microscopic Hamiltonian has a lower symmetry than 
the fixed point Hamiltonian. Any deviations from the full 
symmetry behaviour scale to zero under the renormalization group (RG)
flow when the {\it critical point} is reached. Such phenomenon occurs, for 
instance, in frustrated spin systems near two dimensions.
In the continuum limit, the microscopic Hamiltonian associated 
with Heisenberg spins on the triangular lattice is a non 
linear sigma model with a O(3)$\otimes$ O(2) symmetry\cite{dombre}.
A RG study of this model in $d=2+\epsilon$ dimensions 
show that the symmetry is dynamically enlarged at the 
stable fixed point to O(3)$\otimes$ O(3) $\sim$ O(4)\cite{azaria}.
The phase transition of canted Heisenberg spin systems
belongs thus, near two dimensions,  
to the $N=4$ Wilson-Fisher universality class.

A similar enlarged symmetry scenario has been proposed 
by Zhang in the context of high-T$_c$ cuprates\cite{zhang}. 
The author suggests that the phase diagram of these compounds can 
be deduced from an SO(5) symmetry that unifies antiferromagnetism 
and D-wave superconductivity. It was also argued that 
even though the SO(5) symmetry is only approximate for 
microscopic models such as the Hubbard or t-J models, it 
becomes exact under the RG flow toward a bicritical point. 
Microscopic electron models with an exact SO(5) symmetry
has been constructed\cite{henley} and this symmetry 
occurs in the low energy sector of two-chain Hubbard 
systems\cite{arrigoni,BF}. In particular, Lin, Balents, and 
Fisher\cite{BF}, using a one-loop perturbative RG approach, 
predict that at half filling the weakly-interacting two-leg 
ladder has an exact SO(8) symmetry at low 
energy and belongs to a D-Mott phase where the short range 
pairing correlations have an approximate D-wave symmetry. 
The crucial point of their analysis stems from the fact that some 
coupling constants of the microscopic Hamiltonian 
flow to strong coupling but with ratios converging
to fixed values and the effective model corresponds to 
the SO(8) Gross-Neveu model in the strong coupling limit.

The question that we shall adress, in this letter, 
is whether one can reliably deduce  
restoration of a symmetry by interactions in the strong coupling
limit from weak coupling RG computations. 
In the following, we shall consider several (1+1)-dimensional
models using a non-pertubative approach to investigate
the possible dynamically enlarged symmetry in the 
strong coupling limit.
Let us start with the U(1)-symmetric
Thirring model defined by the following Hamiltonian density: 
\begin{equation}
{\cal H} = \bar{\Psi}_{\alpha} \gamma^{\mu} \partial_{\mu} \Psi_{\alpha}
+ \frac{1}{4} \sum_{a=1}^{3} g_a J^a _{\mu} J^{a \mu}
\label{Thirring}
\end{equation}
with $g_1 = g_2 = g_{\perp}$, and $g_3 = g_{\parallel}$. 
The (iso)spin currents are expressed in terms of the fermion 
fields ($\Psi_{\alpha}$) as:
$J^a _{\mu} = (1/2) \bar{\Psi}_{\alpha} \gamma_{\mu} \sigma^a _{\alpha 
\beta} 
\Psi_{\beta}$, $\sigma^a$ being the Pauli matrices
and our Euclidean conventions for the gamma matrices 
are: $\gamma^0= \sigma^2,\gamma^1= \sigma^1$. 
The Thirring  model (\ref{Thirring}) is 
exactly solvable by the Bethe ansatz 
and the solution
was carefully studied \cite{smirnov,GNW}. 
This gives us the first
opportunity to test whether restoration of symmetry occurs and if
the one-loop approximation in the RG
equations can serve as a reliable tool to detect it.
The RG equations for this model are of the
Kosterlitz-Thouless form: 
\begin{equation}
{\dot g}_{\parallel} = \frac{g^2 _{\perp}}{2\pi},  ~~
{\dot g}_{\perp} = 
\frac{g_{\parallel} g_{\perp}}{2\pi}
\label{RG-XXZ} 
\end{equation}
where the dots indicate derivatives 
with respect to the logarithm of the length scale: ${\dot g} = dg/d\ln
l$.
The phase portrait derived from these equations 
is presented  on Fig. 1; there are regions where 
 the RG trajectories flowing  to strong
coupling converge giving  appearence that the original U(1)
symmetry is enlarged to SU(2). 
The strong coupling regime can be separated into two regions. The
first one corresponds to  $g_{\parallel} >
g_{\perp} > 0$ (marked A on Fig. 1) where the Thirring model 
is equivalent to the sine-Gordon model.  
We shall be interested only in the
region where the symmetry is apparently restored.
In this region the spectrum of the sine-Gordon model
contains only kinks and anti-kinks with equal masses 
and no bound states. All information about  interaction between 
these particles is contained in their two-body S-matrix which depends
explicitly on the renormalized anisotropy: 
\be
\xi = \frac{\pi\mu}{\pi - \mu}, ~~ \mu = \arccos[\cos
 g_{\parallel}/\cos g_{\perp}]  
\approx \sqrt{g_{\parallel}^2 - g_{\perp}^2} \label{anis}
\ee
where the latter equality holds at small bare couplings. 
The S-matrix is an analytic function of $\xi$ at $\xi
\rightarrow 0$ and  corrections to the isotropic limit are order of
$\xi$ at small $\xi$. 
\begin{figure}
	\begin{center}
	\mbox{\psfig{figure=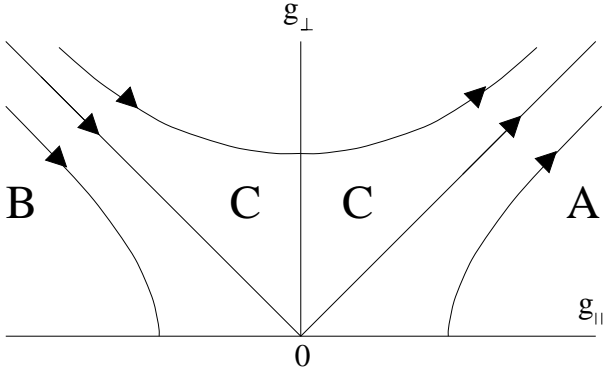,width=8cm}}
	\caption[99]{
	Flow diagram of the U(1) Thirring model.}
	\end{center}
\end{figure}
It turns out, however, that restoration of the 
symmetry is much stronger in the other
region, namely 
$g_{\perp} > |g_{\parallel}|$. Following Japaridze, Nersesyan and
Wiegmann (JNW) who studied it \cite{GNW}, we shall call it region
C (see Fig. 1). The 
JNW solution can be generalized 
for higher representations of the SU(2) group and also for other
semisimple groups\cite{Baz}. 
All these integrable models are deformations of
the Wess-Zumino-Novikov-Witten model by marginally relevant
perturbations. 
Let operators $J^a$ and $\bar J^a$ be the left- and right currents of a
semisimple group ${\cal G}$ satisfying the level $k$ Kac-Moody algebra: 
\be
[J^a(x),J^b(y)] = \frac{\ri k}{4\pi}\delta^{ab}\delta'(x - y) +
\ri f^{abc}J^c(y)\delta(x - y) \label{Kac}
\label{eqcommu}
\ee
with a similar relation for the right spin current ($\bar J$)
whereas $J$ and $\bar J$ commute each other.
In Eq. (\ref{eqcommu}), 
$f^{abc}$ are the structure constant of the algebra
($f^{abc} =
 \epsilon^{abc}$ for the SU(2) group).
The corresponding Hamiltonian density can be written 
in terms of the currents (Sugawara form): 
\bea
{\cal H} = \frac{2\pi}{k + c_v}\sum_{a = 1}^G[:J^aJ^a: + :\bar J^a\bar J^a:] +
  \sum_{a,b=1}^G g_{ab} J^a \bar J^{b}\label{WZW}
\eea
where $G$ is the number of generators of the group $\cal G$ and
the Casimir operator ($c_v$) in the adjoint representation 
is defined by:
$c_v\delta_{a\bar a} = f^{abc}f^{\bar a bc}$. 
 For each ${\cal G}$ and $k$ there is at least one
 {\it single-parametric family} of coupling constants $g^{ab}$ 
(we do not consider  the common  factor as
a parameter) at which the model is integrable. In particular, for
O(2n) groups  there are
two families of solutions.   There is one trouble,
however, -  for all groups higher than SU(2) the parity is broken: 
$g_{ab} \neq g_{ba}$ (to get an idea one may look at 
Refs. \cite{vega,mez,bar} where some
related  Hamiltonians are written explicitly). 
For the SU(2) group there is no trouble and the model (\ref{WZW})
becomes a generalization of the Thirring model with two coupling
constants $g_{\perp}$ and $g_{\parallel}$. At $k = 2$ 
the isotropic SU(2)-invariant version of the model (\ref{WZW}) becomes the
O(3) Gross-Neveu model. This model with $g_{\parallel} =
0$ is equivalent to  the model
suggested by Egger and Gogolin\cite{EG} in the context of theory
of carbon nanotubes:
\be
{\cal H} =  -\frac{\ri}{2}\sum_{a = 1}^3(r_a\p_xr_a - l_a\p_x l_a) +
 g \;r_3l_3 (r_1l_1 + r_2 l_2)
\label{egmodel}
\ee
where $r_a, l_a$ are right and left moving real (Majorana) fermions. 
The equivalence between the SU(2)-invariant version 
model (\ref{WZW}) with $k=2, g_{\parallel} =0$  
and model (\ref{egmodel}) stems from the 
fact that the SU(2)$_2$ spin currents
can be written in terms of three real 
fermions\cite{ZF,shelton}:
$
J^a = -\ri\epsilon^{abc}l_b l_c/2
$
with a similar relation for the right-moving current.
The RG equations of the model (\ref{WZW}) 
in the first loop approximation does not depend on
$k$. Thus in the region of small bare couplings 
the phase portrait of the model is still the one presented on Fig. 1. 

The Bethe ansatz equations for the SU(2)-based  model (\ref{WZW}) 
were obtained by Wiegmann\cite{W}
and  Kirillov and Reshetikhin\cite{KR}
and are given by
\bea
[e_k(x_a + f/2\mu)e_k(x_a - f/2\mu)]^N  = \prod_{b = 1}^Me_2(x_a
- x_b)
\eea
where 
$2N$  is  the number of fermions,  $M$ is the number of up spins.
In the region A the functions $e_n(x)$ are
hyperbolic whereas in the region C they are trigonometric:
\[
e_n(x) = \frac{\sin[\mu(x - \ri n/2)]}{\sin[\mu(x + \ri n/2)]}.
\]
In the region C  the parameters $\mu$ and $0 < f < \pi$ are related 
to the bare coupling
constants (which  are supposed to be small) \cite{GNW}:
\bea
\mu^2 = -g_{\parallel}^2 + g_{\perp}^2, ~~\cot f =
g_{\parallel}/\sqrt{-g_{\parallel}^2 + g_{\perp}^2}.
\eea 
 The results of the previous analysis show that in the region A the
 symmetry is not restored in the model (\ref{WZW}). The region C for the 
case $k > 1$  has never been studied, but it is easy to see that the
 results obtained in Ref. \cite{GNW} for the case $k = 1$ are
 straightforwardly generalizable for greater $k$'s. 
 In the region C the models (\ref{Thirring},\ref{WZW}) 
have the spectral gap 
$
m \approx \Lambda\exp[ - \pi(\pi - f)/2\mu] \label{gap}.
$
 The important result obtained in Ref.\cite{GNW} is that in the strong
coupling limit the anisotropy of the original Hamiltonian leads only
to exponentially small corrections to physical quantities $\sim \exp(-
\pi^2/\mu)$. 
From the technical point of view the most important difference between 
the Bethe ansatz equations in 
the two regions comes from the fact that in the sine-Gordon region
they contain hyperbolic and in the region C - trigonometric
functions. In the latter case the Thermodynamic Bethe Ansatz (TBA)
equations differ from the equations at the isotropic point ($\mu
\rightarrow 0$) only by the fact that real parts of the spectral 
parameters $x_a$
belong to a finite interval $0 < \Re e x < 2\pi/\mu$ and the kernels
in the integral equations are periodic with the period $2\pi/\mu$. As
a matter of fact, one can obtain TBA for the sector C from the
isotropic TBA replacing the kernels by their periodic generalizations:
\be
K_C(x) = \sum_{l = -\infty}^{+\infty}K_{isotr}(x - 2\pi l/\mu).
\ee
At  temperatures comparable with the spectral gap the thermodynamics
is dominated by a small part of the available parameter interval and
the difference between periodic and isotropic kernels becomes
insignificant (exponentially small in $\pi/\mu$, to be precise, 
because kernels in
TBA equations are exponential functions of $x$). 

As we have mentioned, one application of this result is the
Egger-Gogolin model of the carbon nanotubes. In the strong coupling
limit, one can conclude that the 
model (\ref{egmodel}) is equivalent to the O(3) Gross-Neveu model, which
can be described using the results obtained by one
of the authors \cite{Tsv}. There is another
application: the model (\ref{WZW}) with $k \rightarrow \infty$ 
corresponds to 
the model of classical Heisenberg antiferromagnet on a triangular
lattice. At $T = 0$, as consequence of 
frustration,  the system orders in a spiral state with the
order parameter consisting of two mutually perpendicular unit
vectors. In the continuum limit, the long distance properties
are described by a nonlinear sigma model with an order
parameter in SO(3)\cite{dombre,azaria}. 
Using a SU(2) matrix
representation, 
this model is described by the anisotropic 
Principal Chiral (PC) model with the energy  density:
\be
{\cal E} = \frac{\rho_{\perp}}{2}\left[(\Omega_{\mu}^1)^2 +
 (\Omega_{\mu}^2)^2\right] + \frac{\rho_{\parallel}}{2} (\Omega_{\mu}^3)^2
\label{pcmodel}
\ee
where the bare coupling 
constants verify $2\rho_{\perp} = \rho_{\parallel} > 0$.
The vielbein $\Omega_{\mu}^a$ is expressed in terms 
of the matrix $g$ belonging to the SU(2) group by:
$\Omega_{\mu}^a = \ri {\mbox Tr}(\s^a g^{-1}\p_{\mu}g)$.
Since $\rho_{\perp} \ne \rho_{\parallel}$, the bare 
action has a SU(2)$\times$U(1) symmetry.
Using the Polyakov-Wiegmann approach\cite{PW},
one can relate the model (\ref{pcmodel}) to a generalized Thirring model
considered above and thus study the possible dynamically enlarged 
symmetry of the model (\ref{pcmodel}) in the strong coupling limit. 
This correspondence can be established by considering  
the partition function of the anisotropic PC model:  
\be 
Z = \int {\cal D} \Omega_{\mu}^a \; \delta\left(F_{\mu \nu}\right)
\exp \left(-\frac{1}{2} \int d^2 x \; 
\rho_a \Omega_{\mu}^a \Omega_{\mu}^a\right)
\label{partpc}
\ee
where the delta constraint indicates that the gauge field $\sigma^a 
\Omega_{\mu}^a = -2 i g^{-1}\p_{\mu}g$ has a zero field strength.
One has a fermionic representation of this constraint with 
the identity\cite{PW}: 
\bea 
\delta\left(F_{\mu \nu}\right) = \lim_{k\rightarrow +\infty} 
\int {\cal D}{\bar \Psi}_{i\alpha}
{\cal D} \Psi_{i\alpha}
\nonumber \\
\exp \left(-\int d^2x \sum_{i=1}^{k} {\bar \Psi}_{i\alpha}\left(
\gamma^{\mu} \delta_{\alpha \beta} \partial_{\mu} 
+\ri \gamma^{\mu}\Omega_{\mu}^a \sigma^a_{\alpha \beta}\right) \Psi_{i\beta} \right). 
\eea
Inserting this result in (\ref{partpc}) and integrating over the 
$\Omega_{\mu}^a$ fields, one obtains the Hamiltonian 
density of a generalized Thirring model: 
\be
{\cal H} = \sum_{i=1}^{k} 
\bar{\Psi}_{i\alpha} \gamma^{\mu} \partial_{\mu} \Psi_{i\alpha}
+ \frac{1}{2} \sum_{a=1}^{3} {\rho}_a^{-1} J^a _{\mu} J^{a \mu}
\label{Thirringen}
\ee
where the color index $k$ tends to infinity and the current 
is given by: 
$J^a _{\mu} = (1/2) \sum_{i=1}^{k}{\bar\Psi}_{i \alpha} \gamma_{\mu} 
\sigma^a_{\alpha \beta}
\Psi_{i \beta}$. Since the bare coupling constants
of model (\ref{Thirringen}) 
verify ${\rho}_{\perp}^{-1} > {\rho}_{\parallel}^{-1}$,
the anisotropic PC model (\ref{pcmodel}) belongs to the 
region C of the phase portrait of the SU(2)-invariant model (\ref{WZW})
with $k \rightarrow \infty$. 
Therefore, according to the arguments given above,
the anisotropic PC model 
acquires an enlarged SU(2)$\times$SU(2) symmetry in the strong coupling
limit. In this limit its  solution coincides with the
solution of the isotropic Principal Chiral Model (see
Refs. \cite{PW,Sm}). 

The third example we consider is directly related to the work of
Lin, Balents, and Fisher\cite{BF}. It is the anisotropic Gross-Neveu model
with the Hamiltonian density ($U>0$)
\bea
{\cal H} = \bar\psi_0\gamma_{\mu}\p_{\mu}\psi_0 +
\sum_{a = 1}^N\bar\psi_a\gamma_{\mu}\p_{\mu}\psi_a \nonumber\\
-
 g_1(\bar\psi_0\psi_0)^2 + g(\bar\psi_0\psi_0)(\bar\psi_a\psi_a) +
 U(\bar\psi_a\psi_a)^2. \label{GN}
\eea
One can obtain the RG equations 
of the model (\ref{GN}): 
\bea
\dot U &=& -\frac{2(N - 1)}{\pi} \; U^2 - \frac{g^2}{2\pi} \nonumber \\
\dot g &=& -\frac{2N - 1}{\pi} \;gU + \frac{g_1g}{\pi}\nonumber\\
\dot g_1 &=& \frac{N}{2\pi} \;g^2.
\eea
These equations indicate a restoration of the 
symmetry in the strong coupling limit:
the model has an
isotropic behaviour at
strong coupling with $U$ changing its sign and flowing to the
strong coupling. We shall see however, that this isotropy is not
reflected in the excitation spectrum. 

One can, first, bosonize the Dirac fermion $\psi_0$
with the introduction of a bosonic field $\Phi$ to get 
another formulation of (\ref{GN}):
\bea
{\cal H} = \frac{1}{16\pi}(\p_{\mu}\Phi)^2  +
\sum_{a = 1}^N\bar\psi_a\gamma_{\mu}\p_{\mu}\psi_a 
 - \tilde g\cos\beta\Phi(\bar\psi_a\psi_a) \nonumber\\
+
 U(\bar\psi_a\psi_a)^2 \label{GNb}
\eea
where $\tilde g \sim g$ and for $g_1 > 0$
the scaling dimension of the cosine-operator is $d = 2\beta^2 < 1$.
The operator containing the cosine term is then relevant  
and one may think that it will generate a mass gap for all
branches of the spectrum. Thus the mass gaps in this case would be
generated by the repulsive interaction in the 0-channel; this
mechanism was discussed by Shelton and one of the authors in context
of doped ladder models in Ref. \cite{shelton1} where a model very similar to
(\ref{GNb}) with $N = 2$ was considered. 
The model (\ref{GNb}) has O(2)$\times$O(2N) symmetry. 
According to the one-loop RG equations, 
this symmetry is enlarged at strong coupling to
O(2N + 2). 
Let us consider the model (\ref{GNb}) in the large $N$ limit
to test this possible restoration of symmetry. In this
case we can use the mean-field approach to decouple the interaction
between 0 and $a$-sectors:
$
{\cal H} = {\cal H}_{SG} + {\cal H}_{GN}
$
with
\bea
{\cal H}_{SG} = \frac{1}{16\pi}(\p_{\mu}\Phi)^2 - M\cos\beta\Phi\label{sg}\\
{\cal H}_{GN} = \sum_{a = 1}^N\bar\psi_a\gamma_{\mu}\p_{\mu}\psi_a
 - m_0(\bar\psi_a\psi_a) + U(\bar\psi_a\psi_a)^2 \label{gn}
\eea
where $M = \tilde g <(\bar\psi_a\psi_a)>$ and $m_0 = \tilde g <\cos\beta\Phi>$.
Using the exact expression of vacuum expectation values of exponential
fields in the sine-Gordon model\cite{LZ}, one can relate $m_0$ 
to the soliton mass $m_{SG}$ of the sine-Gordon model (\ref{sg}): 
$
m_0 = \tilde g C_d \;m_{SG}^d
$
where the numerical prefactor $C_d$ 
is known\cite{LZ}:
\[
C_d = \frac{\pi\Gamma(1 - d/2)}{2(2 - d)
\sin[\pi d/(2 - d)]\Gamma(d/2)} \sin^d[\pi d/2(2 - d)]
\]
\[
\times\left\{\frac{2\sqrt\pi}{\Gamma[1/(2 - d)]
\Gamma[1 - d/2(2 - d)]}\right\}^{2 - d}.
\]
Using the large $N$ limit of model (\ref{gn}), one obtains 
the following expressions:
\bea
M = \frac{\tilde g N}{2\pi}m\ln(\Lambda/m)\\
m = \frac{m_0}{1 + [2UN/\pi]\ln(\Lambda/m)}.
\eea
In order to close the different equations
of the large $N$ limit of the model (\ref{GNb}), we need a relation
between $M$ and $m_{SG}$ which is given by:
$M = D_d m_{SG}^{2 - d}$
with\cite{LZ}
\[
D_d = \frac{2\Gamma(d/2)}{\pi\Gamma(1 - d/2)}\left\{\frac{\sqrt{\pi}
\Gamma[1/(2 - d)]}{2\Gamma[d/2(2 - d)]}\right\}^{2 - d}.
\]
We shall assume that in the limit $N \rightarrow \infty$,  
$UN = U^*, {\tilde g}^2N^d = g^*$ remain finite and that
$U^*\ln(\Lambda/m) >> 1$. In that case, we obtain  
the mass spectrum of the model in the large $N$ limit:
\bea
m_{SG} &\simeq& \left(\frac{{\tilde g}^2N}{4U^*}\frac{C_d}{D_d}\right)^{1/(2 -
2d)}\nonumber\\
m\ln(\Lambda/m) &\simeq& 2\pi D_d (g^*)^{1/(2 - 2d)}(C_d/4D_d U^*)^{(2-
d)/(2 - 2d)} \nonumber.
\eea
As wee see, the ratio of the masses $m_{SG}/m \sim \sqrt N$ is large.
Therefore, this result indicates 
that the symmetry is not restored by interactions in the 
strong coupling limit. 

From our discussion of generalized Thirring models,
we conclude that there are two possible sectors
where the symmetry is apparently restored in the 
strong coupling regime. In the first one, called the
sine-Gordon sector, the parameter characterizing 
anisotropy does not renormalize to zero
at strong coupling. Its value depends on the
bare coupling constants and is unlikely to be small outside the region
of weak bare interactions. There is another region, however, the
region C, or the trigonometric region where the influence of the 
anisotropy is exponentially small. Here the restored symmetry 
 has more chances to survive outside of regions of weak bare
couplings. Both regions are indistinguishable
within the perturbative RG approach in the infrared limit. 
We have also given an example
where the RG equations indicate isotropy,
but the nonperturbative approach based on large N-approximation shows stricking
anisotropy.

 We are  grateful to F. Essler, A. Gogolin, A. Nersesyan, and S. Sachdev for
 interesting discussions and interest to the work.

\end{document}